\documentclass[aps,pre,twocolumn,superscriptaddress,fleqn,tbtags,onecolumn
]{revtex4}%
\usepackage{pdfpages}
\usepackage{listings}
\usepackage[english]{babel}%
\usepackage{amsmath,amssymb}%
\usepackage{leftidx}%
\usepackage{graphicx}%
\usepackage{grffile}%
\usepackage{ifthen}%
\usepackage[squaren]{SIunits}%
\usepackage{xcolor}%
\usepackage{hyperref}%
\usepackage{dcolumn}
\usepackage{bm}
\usepackage{xr}
\usepackage{url}

\usepackage{ulem}

\newcommand{\cf}{{cf.\ }}                            
\newcommand{\eq}{eq.}                                     
\newcommand{\Eq}{Eq.}                                     
\newcommand{\eqs}{eqs.}                                   
\newcommand{\file}[1]{\nolinkurl{#1}}                           
\newcommand{\fig}{fig.}                                   
\newcommand{\figs}{figs.}                                 
\newcommand{\ie}{{i.\,e.\ }}                         
\newcommand{\eg}{{e.\,g.\ }}                         

\newcommand{\via}{{via}\ }                             

\newcommand{\treref}[2]{[notes p.~#1 %
  \ifthenelse{\equal{#2}{}}{} {eq.~(#2)}%
]}                                                  
\newcommand{\trerevisiondelete}[1]{\textbf{\sout{#1}}}
\renewcommand{\trerevisiondelete}[1]{}

\newcommand{\order}{\mathcal O}
\newcommand{\BesselKsymbol}{\mathrm{K}}
\newcommand{\BesselK}[2]{\BesselKsymbol_{#1}(#2)}

\newcommand{\eqindent}{\quad}                             
\newcommand{\eqspace}{\;}                              
\newcommand{\LangevinFunction}{\mathcal L}                
\newcommand{\matsc}[3]{#1 
  \ifthenelse{\equal{#2}{}}{}{^{(#2)}}%
  \ifthenelse{\equal{#3}{}}{}{_{#3}}%
}%
\newcommand{\vecsc}[3]{#1 
  \ifthenelse{\equal{#2}{}}{}{^{(#2)}}%
  \ifthenelse{\equal{#3}{}}{}{_{#3}}%
}%
\addunit{\molar}{M}                                       
\addunit{\calory}{cal}                                    

\newcommand{\avogadro}{\mathrm{N_A}}           

\newcommand{\con}{\rho}                                      
\newcommand{\conA}{\con_{\mathrm{a}}}                     
\newcommand{\conK}{\con_{\mathrm{c}}}                     
\newcommand{\drm}{\mathrm{d}}                  
\newcommand{\kB}{\mathrm{k_B}}                            
\newcommand{\kBT}{\mathrm{k_B}T}                          
\newcommand{\dielectricO}{\varepsilon_0}                   
\newcommand{\dielectricr}{\varepsilon_{\mathrm{r}}}                   
\newcommand{\valence}{z}                                              
\newcommand{\valenceA}{\valence_{\mathrm{a}}}                     
\newcommand{\valenceK}{\valence_{\mathrm{c}}}                     

\newcommand{\Rbp}[2]{(#1,#2)}                             
\newcommand{\Rc}{c}                                       
\newcommand{\RF}{F}                                       
\newcommand{\RG}{\mathcal G}                              
\newcommand{\Rg}{ g}                                      
\newcommand{\RGf}{\RG_{\mathrm{f}}}                       
\newcommand{\Rgf}{\Rg_{\mathrm{f}}}                       
\newcommand{\RGfFJC}{\RGf^{\mathrm{FJC}}}
\newcommand{\RgfFJC}{\Rgf^{\mathrm{FJC}}}
\newcommand{\RGh}{\RG_{\mathrm{h}}}                       
\newcommand{\RGhstack}{\RG_{\mathrm{h}}^{\mathrm{stack}}} 
\newcommand{\RGhsalt}{\RG_{\mathrm{h}}^{\mathrm{salt}}}   
\newcommand{\Rghelectrostatic}{\Rg_{\mathrm{h}}^{\mathrm{DH}}}   
\newcommand{\RGhinit}{\RG_{\mathrm{h}}^{\mathrm{init}}}   
\newcommand{\RGhterm}{\RG_{\mathrm{h}}^{\mathrm{term}}}   
\newcommand{\Rgh}{\Rg_{\mathrm{h}}}                
\newcommand{\Rghstack}{\Rgh^{\mathrm{stack}}}                
\newcommand{\RGl}{\RG_{\mathrm{l}}}                   
\newcommand{\RGlentropy}{\RG_{\mathrm{l}}^{\mathrm{conf}}} 
\newcommand{\RGlinit}{\RG_{\mathrm{l}}^{\mathrm{init}}} 
\newcommand{\RGlsalt}{\RG_{\mathrm{l}}^{\mathrm{salt}}} 
\newcommand{\Rheatcap}{C}                       
\newcommand{\Rionicstrength}{I}                              
\newcommand{\Rkappa}{\kappa}                              
\newcommand{\RKuhnss}{b_{\mathrm{ss}}}                    
\newcommand{\RlB}{l_{\mathrm{B}}}                         
\newcommand{\Rlss}{l_{\mathrm{ss}}}                       
\newcommand{\Rlds}{l_{\mathrm{ds}}}                       
\newcommand{\Rlinecharge}{\tau}                           
\newcommand{\Rlinechargess}{\Rlinecharge_{\mathrm{ss}}}   
\newcommand{\Rlinechargeds}{\Rlinecharge_{\mathrm{ds}}}   
\newcommand{\RM}{M}                         
\newcommand{\Rm}{m}                         
\newcommand{\RN}{N}                         
\newcommand{\Runitcharge}{e_0}                            
\newcommand{\Rvf}[1]{v_{\mathrm{f}}(#1)}                  
\newcommand{\RT}{T}                                       
\newcommand{\RTmelt}{\RT_{\mathrm{m}}}                    
\newcommand{\RQH}{Q}               
\newcommand{\RQHt}{\tilde Q}   
\newcommand{\RQG}{Z}              
\newcommand{\subref}[2]{(#2)}
\newcommand{\subfloat}[1]{}
\begin{document}

\title{ Theory for RNA folding, stretching, and melting including
  loops and salt} \author{Thomas R. Einert} \email[Corresponding
author:]{ Physik Department, Technische Universit\"at M\"unchen,
  James-Franck-Stra\ss e, 85748 Garching, Germany, Tel.:
  +49-89-28914337, Fax: +49-89-28914642, E-mail: einert@ph.tum.de}
\affiliation{Physik Department, Technische Universit\"at M\"unchen,
  85748 Garching, Germany}%
\author{Roland R. Netz} \email[E-mail: ]{netz@ph.tum.de}
\affiliation{Physik Department, Technische Universit\"at M\"unchen,
  85748 Garching, Germany}

\date{\today}
\begin{abstract}
  Secondary structure formation of nucleic acids strongly depends on
  salt concentration and temperature.  We develop a theory for RNA
  folding that correctly accounts for sequence effects, the entropic
  contributions associated with loop formation, and salt effects.
  Using an iterative expression for the partition function that
  neglects pseudoknots, we calculate folding free energies and minimum
  free energy configurations based on the experimentally derived base
  pairing free energies.
 The configurational entropy of loop formation is modeled by the
  asymptotic expression $ -\Rc \ln \Rm$, where $\Rm$ is the length of
  the loop and $\Rc$ the loop exponent, which is an adjustable
  constant.  Salt effects enter in two ways: first, we derive salt
  induced modifications of the
 free energy parameters for describing base pairing and, second, we
  include the electrostatic free energy for loop formation.  Both
  effects are modeled on the Debye-H\"uckel level including counterion
  condensation.  We validate our theory for two different RNA
  sequences: For tRNA-phe, the resultant heat capacity curves for
  thermal denaturation at various salt concentrations accurately
  reproduce experimental results.  For the P5ab RNA hairpin, we derive
  the global phase diagram in the three-dimensional space spanned by
  temperature, stretching force, and salt
 concentration 
and obtain good agreement with the experimentally determined 
  critical unfolding force.
  We show that for a proper description of RNA melting and stretching,
  both salt and loop entropy effects are needed.
\end{abstract}
\keywords{RNA, DNA, force spectroscopy, melting, salt dependence, loop
  entropy}

\maketitle


\section{Introduction}
\label{sec:introduction}

Ribonucleic acid (RNA) is one of the key players in molecular biology
and has in the past attracted theoretical and experimental physicists
because of its intriguing structural and functional properties. RNA
has multiple functions: beyond being an information carrier it has
regulatory and catalytic abilities~\cite{Gesteland2005}. Comprehending
how RNA folds and what influences the folding process are key
questions~\cite{Tinoco1971}. Thus, the reliable prediction of RNA
structure and stability under various conditions is crucial for our
understanding of the functioning of RNA and nucleic acid constructs in
general~\cite{Liedl2007,Dietz2009}.

The influence of temperature and solution conditions on RNA folding
stays in the interest of experimental groups. Traditionally the
thermal melting of RNA was monitored \via differential scanning
calometry or UV spectroscopy for the bulk
ensemble~\cite{Xia1998,Mathews1999,Privalov1978,Vives2002}. More
recently, single molecule pulling and unzipping experiments have been
used to unveil the influence of different solution conditions and even
determine energy
parameters~\cite{Liphardt2001,Vieregg2007,Huguet2010}.

On the theoretical side, RNA denaturation has been modeled on various
levels of coarse graining. Focusing on the secondary structure, namely
the base pairs~(bp), and omitting tertiary interactions, equilibrium
folding and unfolding has been modeled very
successfully~\cite{McCaskill1990,Zuker1981,Markham2005,Gerland2001,Bundschuh2005,Montanari2001,Mueller2003,Dimitrov2004,Imparato2009a,Einert2008,Einert2011a}. In
the presence of a logarithmic contribution to the loop entropy, it has
been shown that homopolymeric RNA, where sequence effects are
neglected, features a genuine phase transition, which can be induced
by force or
temperature~\cite{Mueller2002,Mueller2003,Einert2008,Einert2011a}.
However, the specific sequence influences the stretching response of a
molecule, which has been shown by \citet{Gerland2001,Gerland2003}, yet
without considering the logarithmic loop entropy. More detailed
insights can be obtained by simulations, which are numerically quite
costly, though, when compared to models focusing only on secondary
structure.  Coarse grained, Go-like simulations of short RNA hairpins
allowed to analyze the dynamics of the folding and unfolding
process~\cite{Hyeon2005,Hyeon2006}.  Ion specific effects have been
studied by performing molecular dynamics~\cite{Auffinger2000} or
coarse grained simulations~\cite{Tan2006,Tan2007,Tan2008, Jost2009}.
Much less is known about the salt dependence of denaturation
transitions of RNA.

While for DNA numerous corrections of the base pairing free energies
due to varying salt concentration exist, see~\cite{Owczarzy2004} and
references therein, analogous results for the salt dependence of RNA
energy parameters are sparse~\cite{Tan2006}. However, molecular
biology and biotechnological applications depend on the reliable
prediction of RNA stability for different solution conditions.

In this paper we extend these previous works and develop a theory that
allows to include all these effects --~sequence, salt dependence,
logarithmic loop entropy, stretching force~-- and demonstrate that all
are necessary to obtain a complete picture of the thermodynamics of
the secondary structure of RNA.
Neglecting tertiary interactions, we use a recursion relation, which
allows to correctly account for logarithmic and thus non-linear free
energy contributions due to the configurational entropy of
loops~\cite{Einert2008}. To include the influence of monovalent salt
on RNA stability, we model the RNA backbone as a charged polymer
interacting \via a Debye-H\"uckel potential and give heuristic
formulas for the modification of the loop free energy and the base
pairing and stacking free energy parameters.
Debye-H\"uckel is a linear theory, yet we include non-linear effects
caused by counterion condensation using Manning's
concept~\cite{Manning1969}.  The backbone elasticity of single
stranded RNA (ssRNA) is described by the freely jointed chain (FJC)
model. Our description allows for a complete description of the
behavior of RNA in the three-dimensional phase space spanned by
temperature, salt concentration, and external stretching force. We
find that for an improved description of RNA melting curves one needs
to include both salt effects and loop entropy. Only the combined usage
of these two contributions enables to predict the shift of the melting
temperature (due to salt) and the cooperativity (due to logarithmic
loop entropy), which is illustrated in the case of
tRNA-phe. As an independent check we consider the
  force induced unfolding of the P5ab RNA hairpin and observe
good agreement with experimental values with no fitting
  parameters.  The influence of salt is
illustrated by melting curves and force extension curves for various
salt concentrations. For the P5ab hairpin the phase diagram is
determined and slices through the three-dimensional parameter space are
shown.

\section{Free energy parameterization}

RNA folding can be separated into three steps, which occur
subsequently and do not influence each other to a fairly good
approximation~\cite{Tinoco1999}. The primary structure of RNA is the
mere sequence of its four bases cytosine~(C), guanine~(G),
adenine~(A), and uracil~(U). Due to base pairing, \ie either the
specific interaction of C with G or the interaction of A with U, the
secondary structure is formed. Therefore, on an abstract level, the
secondary structure is given by the list of all base pairs present in
the molecule. Only after the secondary structure has formed, tertiary
contacts arise. Pseudoknots~\cite{Richards1969,Batenburg2000}, helix
stacking, and base triples~\cite{Higgs2000} as well as the overall
three-dimensional arrangement of the molecule are considered as parts
of the tertiary structure. The main assumption of hierarchical folding
is, that tertiary structure formation operates only on already
existing secondary structure elements~\cite{Tinoco1999}.  Although
cases are known where this approximation breaks down, it generally
constitutes a valid starting point~\cite{Cho2009}.  In this paper,
where the main point is the influence of the loop entropy and the salt
concentration on the secondary structure, we therefore neglect
tertiary interactions altogether.
\begin{figure}
  \centering
  \includegraphics{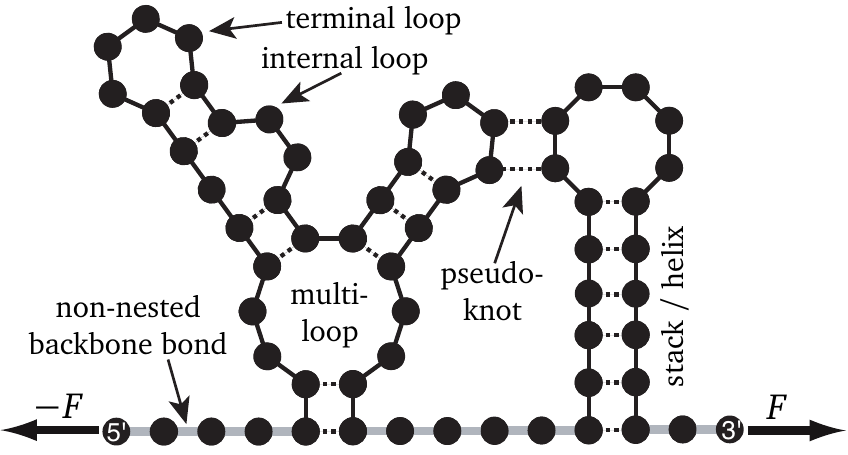}
  \caption{Schematic representation of the secondary structure of an
  RNA molecule. Dots represent one base, \ie cytosine, guanine,
  adenine, or uracil. Solid lines denote the sugar-phosphate
  backbone bonds, broken lines base pairs, and thick gray lines the
  non-nested backbone bonds, which are counted by the variable
  $\RM$, here $\RM=11$. The thick arrows to either side illustrate
  the force~$\RF$ applied to the 5'- and 3'-end.}
  \label{fig:1}
\end{figure}

Given a set of base pairs, the secondary structure consists of helices
and loops as the basic structural units, \cf \fig~\ref{fig:1}.
Since pseudoknots are neglected, every nucleotide can be attributed
unambiguously to exactly one subunit. The free energy of a certain
secondary structure is then given by the sum of the free energy
contributions of the individual structural subunits, as we will detail
now.

\subsection{Free energy of a loop}
\label{sec:free-energy-loop}

We model the free energy of a loop consisting of $m$ backbone bonds,
see \fig~\ref{fig:1}, with
\begin{equation}
  \label{eq:1}
  \RGl (\Rm)  = \RGlentropy (\Rm) + \RGlsalt (\Rm) + \RGlinit
  \eqspace.
\end{equation}
The first term is the loop entropy difference between an unconstrained
polymer and a ring-like polymer, which is characterized by the loop
exponent~$\Rc$~\cite{Duplantier1986,Kafri2000,Einert2008,Mueller2002}
\begin{equation}
  \label{eq:2}
  \RGlentropy (\Rm) = - \kBT \ln \Rm^{-\Rc}
  \eqspace,
\end{equation}
with $\kB$ the Boltzmann constant and $\RT$ the absolute temperature.
The loop exponent~$\Rc$ is $\Rc_{\mathrm{ideal}}=3/2$ for an ideal
polymer and $\Rc_{\mathrm{SAW}} = 1.76$ for an isolated self avoiding
loop. Helices emerging from the loop limit the configurational space
available to the loop and hence increase $\Rc$. One obtains
$\Rc_1=2.06$ for terminal, $\Rc_2=2.14$ for internal loops and
$\Rc_4=2.16$ for a loop with four emerging helices~\cite{Einert2008}.
Since the differences between these exponent values are quite small,
we assume a constant loop exponent $\Rc = 2.1$ in this paper and only
compare with the case of vanishing loop entropy characterized by
$c=0$.

The second term in \eq~\eqref{eq:1} describes the free energy
difference between a charged ring of length $m\Rlss$ and a straight
rod of the same length due to electrostatic interactions, with $\Rlss
= \unit{6.4}{\angstrom}$ the length of one ssRNA backbone
bond~\cite{Tan2008}.  The electrostatics are modeled on the
Debye-H\"uckel level~\cite{Kunze2002}
\begin{align}
  \label{eq:3}
  \RGlsalt (\Rm) &= \kBT \RlB (\Rm \Rlss) \Rlinechargess^2 \Biggl[
  \ln(\Rkappa \Rm\Rlss) - \ln(\pi/2) + \gamma
  - \frac{\Rkappa \Rm\Rlss}{2} \leftidx{_1}{F}{_2}\left(
    1/2, \begin{pmatrix} 1\\3/2 \end{pmatrix}, \left(\frac{\Rkappa
        \Rm\Rlss}{2\pi}\right)^2
  \right)\notag\\
  & \eqindent\eqindent + \frac12 \left(\frac{\Rkappa
      \Rm\Rlss}{\pi}\right)^2 \leftidx{_2}{F}{_3}\left(
    \begin{pmatrix}
      1\\1
    \end{pmatrix},
    \begin{pmatrix}
      3/2\\3/2\\2
    \end{pmatrix}, \left(\frac{\Rkappa \Rm\Rlss}{2\pi}\right)^2
  \right) \notag\\ & \eqindent\eqindent + \frac{1}{\Rkappa
    \Rm\Rlss}\left(1-\exp(-\Rkappa \Rm\Rlss) + \Rkappa \Rm\Rlss
    \Gamma(0, \Rkappa \Rm\Rlss)\right) \Biggr] \eqspace,
\end{align}
with $\RlB = \Runitcharge^2/(\kBT 4\pi\dielectricO\dielectricr) $ the
Bjerrum length, which in water has a value of roughly
$\unit{7}{\angstrom}$, $\Rkappa^{-1} =
\sqrt{\dielectricO\dielectricr\kBT/(2\avogadro\Runitcharge^2I) }$ the
Debye screening length, $\dielectricO$ the vacuum dielectric constant,
$\dielectricr\approx80$ the relative dielectric constant of
water~\cite{Murrell1994}, $\Rionicstrength = 1/2(\conA \valenceA^2 +
\conK\valenceK^2)$ the ionic strength, $\conA/\conK$ and
$\valenceA/\valenceK$ the concentration and the valency of the
anions/cations, $\avogadro$ the Avogadro constant, $\Runitcharge$ the
elementary charge, $\gamma\approx0.58$ Euler's constant, $\Gamma(a,x)$
the incomplete gamma function, and $\leftidx{_p}{F}{_q}$ the
generalized hypergeometric functions~\cite{Abramowitz2002}. To account
for modifications of the line charge density $\Rlinechargess $ 
due to non-linear electrostatic effects,
we employ Manning's counterion condensation theory~\cite{Manning1969},
predicting
\begin{equation}
 \Rlinechargess = \min(1/ \Rlss,  1/(\RlB\valenceK))\label{eq:4}
  \eqspace.
\end{equation}
\Eq~\eqref{eq:3} amounts to a ground state approximation of the
electrostatic contribution to the free energy of a loop. This is
rationalized by the fact that the electrostatic interaction is
screened and decays exponentially over the Debye length, which is
roughly $\Rkappa^{-1} = \unit{1}{\nano\meter}$ for
\unit{100}{\milli\molar} salt solution.  However, typical distances
between bases in a loop are of the order of the helix diameter
$d=\unit{2}{\nano\meter}$ or larger. Therefore, we expect
electrostatic interactions to be basically independent of the global
configuration of a loop, which justifies both the ground state
approximation and our additivity approximation, where ion effects and
conformational contributions decouple, see \eq~\eqref{eq:1}.  In the
supporting material, see \eq~S4, we give
an interpolation formula for \eq~\eqref{eq:3} involving no
hypergeometrical functions.

\begin{figure}
  \centering
  \subfloat{\label{fig:2a}}%
  \subfloat{\label{fig:2b}}%
  \includegraphics{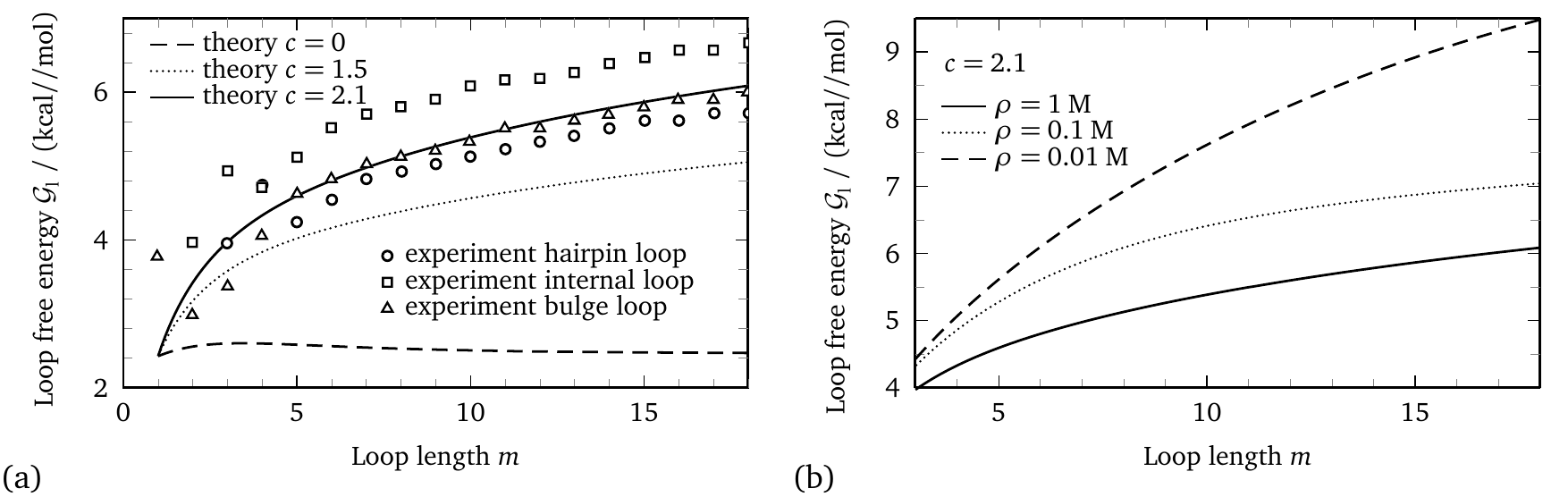}
  \caption{\subref{fig:2}a~Free energy of a loop as a function of the number of segments $m$
    for different loop exponents $\Rc = 0,\,1.5,\,2.1$ (lines) and for NaCl concentration $\con
    =\unit{1}{\molar} $.
Symbols denote experimental values for various types of loops
(hairpin, bulge, internal)~\cite{Serra1995,Tan2008} for $\con =
\unit{1}{\molar}$ NaCl.  $\RGlinit$ is obtained by fitting
$\RGl(\Rm)$, \eq~\eqref{eq:1}, to the experimental data for $\Rc =
2.1$ and $\con = \unit{1}{\molar}$.  The same salt concentration and
the same value for $\RGlinit$ is used for plotting the curves with
$\Rc = 0,\,1.5$.  \subref{fig:2}b~Salt dependence of the free energy
of loops as a function of the number of segments $m$ for different
salt concentrations $\con = \unit{1}{\molar},\
\unit{0.1}{\molar},\,\unit{0.01}{\molar}$ and loop exponent $\Rc =
2.1$ according to \eq~\eqref{eq:1}.  }
\label{fig:2}
\end{figure}

The last term in \eq~\eqref{eq:1} is the loop initiation free energy
$\RGlinit$.  As we are employing a logarithmic loop entropy,
\eq~\eqref{eq:2}, we cannot use the standard value for $\RGlinit$,
which was extracted from experimental data  for a different loop
parameterization~\cite{Xia1998,Mathews1999}.
Therefore, a modified  value $\RGlinit$ is obtained by fitting $\RGl(\Rm)$,
given by \eq~\eqref{eq:1}, to experimental data using $\Rc = 2.1$ in
$\RGlentropy(\Rm)$ and the salt concentration $\con =
\unit{1}{\molar}$ in $\RGlsalt(\Rm)$, see \fig~\ref{fig:2}a.  In this
figure we show experimentally determined free energies for terminal,
internal and bulge loops as a function of the loop size,
which exhibit a dependence on the type of the loop.
As an approximation, we do not
distinguish between those loop types in the theory and consequently
fit a single parameter $\RGlinit$ to the data, which turns out to be
$\RGlinit= \unit{1.9}{\kilo\calory\per\mole}$ for $\RT =
\unit{300}{\kelvin}$, see supporting material
section~C.
In \fig~\ref{fig:2}a the fitted $\RGl(\Rm)$ for the loop exponent
$\Rc=2.1 $ is depicted by the solid line; the other lines illustrate
the effect of different loop exponents on the loop free energy
according to \eq~\eqref{eq:1} using the same value for $\RGlinit$.
\fig~\ref{fig:2}b illustrates the effect of salt on the loop free energy for a given 
value of $\Rc=2.1$.

\subsection{Free energy of a helix}
\label{sec:free-energy-helix}

The free energy of a helix
\begin{equation}
  \label{eq:5}
  \RGh = \RGhstack + \RGhinit + \RGhterm + \RGhsalt 
  \eqspace
\end{equation}
depends on the sequence $\{b_i\}$, which consists of the four
nucleotides $b_i =\mathrm{C,G,A,U}$.  The stacking free
energy~$\RGhstack$ is based on experimentally determined parameters
incorporating the base pairing free energy as well as the stacking
free energy between neighboring base pairs. In the standard notation,
$\Rghstack[\Rbp{b_{i}}{b_{j}},\Rbp{b_{i+1}}{b_{j-1}}]$ is the
contribution of the two neighboring, stacked base pairs
$\Rbp{b_{i}}{b_{j}}$ and $\Rbp{b_{i+1}}{b_{j-1}}$ to $\RGhstack$. The
explicit values for the enthalpic and entropic parts are given in the
supporting material.  We use the expanded nearest neighbor
model~\cite{Mathews1999,Xia1998} to calculate the base pairing and
stacking contributions of a helical section ranging from base pair
$\Rbp{i}{j}$ through $\Rbp{i+h}{j-h}$ and obtain
\begin{equation}
  \label{eq:6}
  \RGhstack = \sum_{h'=1}^h\Rghstack[(b_{i+h'-1}, b_{j-h'+1}),(b_{i+h'}, b_{j-h'})]
  \eqspace.
\end{equation}
The initiation and termination free energies in \eq~\eqref{eq:5} take
into account weaker pairing energies of AU or GU base pairs at the
ends of the helix. We use the standard literature values for
$\RGhinit$ and $\RGhterm$~\cite{Mathews1999,Xia1998} and summarize the
explicit values in the supporting material. Increasing the salt
concentration increases the stability of a helix: First, counterions
condense on the negatively charged backbone and reduce the
electrostatic repulsion and, second, the diffuse counterion cloud
surrounding the charged molecule screens the interaction. We model the
two strands of a helix as two parallel rods at distance
$d=\unit{2}{\nano\meter}$ interacting \via a Debye-H\"uckel potential
characterized by the screening length $\Rkappa^{-1}$.  The
electrostatic interaction energy per nucleotide with the other strand
is given by
\begin{equation}
  \label{eq:7}
  \Rghelectrostatic (\con)= \kBT\Rlinechargeds^2\Rlds^{}\RlB^{}
  \int_{-\infty}^\infty \frac{\exp(-\Rkappa\sqrt{d^2+z^2})}{\sqrt{d^2+z^2}} dz   =
  2 \kBT\Rlinechargeds^2\Rlds^{}\RlB^{} \BesselK{0}{\Rkappa d}
  \eqspace.
\end{equation}
$\Rlds = \unit{3.4}{\angstrom}$ is the helical rise per base pair of
double-stranded RNA (dsRNA) and $\BesselK{0}{\Rkappa d}$ is the zeroth
order modified Bessel function of the second kind.  Again, we employ
Manning's theory~\cite{Manning1969} to calculate the line charge
density $\Rlinechargeds  = \min(1/  \Rlds, 1/(\RlB\valenceK))$.  The reference
state for the salt correction of the pairing free energy is at
temperature $\RT = \unit{300}{\kelvin}$ with monovalent salt
concentration $\con = \unit1\molar$, as the experimental pairing free
energies~$\Rghstack$ were determined at this concentration.
The free energy shift for a helix consisting of $h$ base pairs due to
electrostatic interactions is then
\begin{equation}
  \label{eq:8}
  \RGhsalt = h( \Rghelectrostatic (\con) -   \Rghelectrostatic (\unit{1}{\molar}))
  \eqspace.
\end{equation}

The use of Debye-H\"uckel theory to incorporate salt effects enables
to include the overall dependence on temperature and salt
concentration but involves several approximations. First, we are using
Manning's counterion condensation theory to obtain the actual line
charge density of ssRNA and dsRNA~\cite{Manning1969}. However, Manning
condensation is known to underestimate the line charge at increasing
salt concentration and therefore favors the bound
state~\cite{Netz2003}.  Second, when calculating the electrostatic
energy of a loop we effectively use a ground state approximation and
neglect conformational fluctuation effects.
Third, when two ssRNA strands come together to form a helix, the line
charge density increases since the distance between two bases
decreases. The salt dependence of the work to decrease the axial distance
between two bases from $\Rlss = \unit{6.4}{\angstrom}$ to $\Rlds =
\unit{3.4}{\angstrom}$ is neglected. This approximation favors the
unbound state.  Therefore, it is very important to validate the model
we employ, which we do by detailed comparison with experimental
data. From the favorable comparison with experiments we tentatively
conclude that the various errors partially cancel and the resulting
expression for the salt influence is quite accurate.
We point out that after determining $\RGlinit$ in
\eq~\eqref{eq:1}, no further fitting is done and only standard
literature values are used.

Our theory is able to consider variations of the salt concentration as
well as of the temperature, which makes it suitable to study RNA
melting at various salt concentrations in a consistent way. However,
since our approach is solely based on mean field theory, it will become unreliable
in the case of multivalent ions, where correlations become
important. Also, ion specific effects, which are important for
divalent ions such as Mg\textsuperscript{2+}~\cite{Draper2005}, are
not considered in our approach.

\subsection{Response of the molecule to an external stretching force}
\label{sec:response-molecule-an}
In atomic force microscope or optical tweezers experiments, it is
possible to apply a stretching force $\RF$ to the two terminal bases
of the molecule. We model the stretching response of the $\RM$
non-nested backbone bonds, see \fig~\ref{fig:1}, with the freely
jointed chain (FJC) model
\cite{Montanari2001,Gerland2001,Gerland2003}. A non-nested bond is
defined as a backbone bond, which is neither part of a helix nor part
of a loop. It is outside all secondary structure elements and
therefore contributes to the end-to-end extension observed in force spectroscopy
experiments.  The force dependent contribution to the free energy per
non-nested monomer is given by
\begin{equation}
  \label{eq:9}
  \RgfFJC = \RGfFJC / \RM = -\kBT \frac{\Rlss}{\RKuhnss} \ln
  \left(\frac{\sinh(\beta\RKuhnss\RF)}{\beta\RKuhnss\RF}\right)
  \eqspace,
\end{equation}
where $\beta = 1/(\kBT)$ is the inverse thermal energy and $\RKuhnss =
\unit{1.9}{\nano\meter}$ is the Kuhn length of
ssRNA~\cite{Montanari2001} (we used the Kuhn length of ssDNA as the
corresponding ssRNA data is less certain).  The stretching response of
one non-nested monomer to an external force is then given by
\begin{equation}
  \label{eq:10}
  x^{\mathrm{FJC}}(\RF) = -\frac{\drm\RgfFJC}{\drm\RF}  = \Rlss \LangevinFunction (\beta\RF \RKuhnss) =
  \Rlss\left( \coth(\beta\RF \RKuhnss) + 1/ (\beta\RF \RKuhnss)\right)
  \eqspace,
\end{equation}
$\LangevinFunction$ is the Langevin function. Electrostatic effects on
the stretching response are considered to be small and hence are
neglected~\cite{Marko1995,Netz2001a}.

\section{Calculation of the partition function}
\label{sec:calc-part-funct}

So far we showed how to calculate the free energy of one given
secondary structure. The next step is to enumerate all possible
secondary structures and to obtain the partition function, which
allows to study the thermodynamics of the system.  As we neglect
tertiary contacts --~and in particular pseudoknots~-- for any two base
pairs $\Rbp{i}{j}$ and $\Rbp{k}{l}$ with $i<j$, $k<l$, and $i<k$ we
have either $i<k<l<j$ or $i<j<k<l$. This allows to derive a recursion
relation for the partition function of the secondary structure.  In
our notation, the canonical partition function $\RQH_{i,j}^{\RM}$ of a
sub-strand from base $i$ at the 5'-end through $j$ at the 3'-end
depends on the number of non-nested backbone bonds
$\RM$~\cite{Einert2008,Bundschuh2005,Mueller2002}, see
\fig~\ref{fig:1}.  The recursion relations for $\RQH_{i,j}^{\RM}$ can
be written as
\begin{subequations}\label{eq:11}
  \begin{equation}
    \label{eq:9a}
    \RQH_{i,j+1}^{\RM+1} = \frac{\Rvf{\RM+1}}{\Rvf{\RM}} \left[ \RQH_{i,j}^{\RM} +
      \sum_{k=i+\RM+1}^{j-N_{\mathrm{loop}}}\RQH_{i,k-1}^{\RM} \RQH_{k, j+1}^{0}\right]
  \end{equation}
  and
  \begin{equation}
    \label{eq:9b}
    \RQH_{k,j+1}^{0}= \sum_{h=1}^{(j-k-N_{\mathrm{loop}})/2}
    \exp[-\beta{\RGh}^{\Rbp{k}{j+1}}_{\Rbp{k+h}{j+1-h}}]\sum_{m=1}^{j-k-1-2h}
    \RQH_{k+1+h,j-h}^{m}\frac{\exp[-\beta\RGl(m+2)]}{\Rvf{m}}
    \eqspace.
  \end{equation}
\end{subequations}
\Eq~\eqref{eq:9a} describes elongation of an RNA structure by either
adding an unpaired base (first term) or by adding an arbitrary
sub-strand $ \RQH_{k, j+1}^{0}$ that is terminated by a helix.
\Eq~\eqref{eq:9b} constructs $ \RQH_{k, j+1}^{0}$ by closing
structures with $\Rm$ non-nested bonds, summed up in $
\RQH_{k+1+h,j-h}^{\Rm}$, by a helix of length $h$.
$N_{\mathrm{loop}}=3$ is the minimum number of bases in a terminal
loop.  $\Rvf{\RM}$ denotes the number of configurations of a free
chain with $\RM$ links and drops out by introducing the rescaled
partition function $\RQHt_{i,j}^{\RM}=\RQH_{i,j}^{\RM}/\Rvf{\RM}$ and
will not be considered further since its effects on the partition function are negligible.
${\RGh}^{\Rbp{k}{j+1}}_{\Rbp{k+h}{j+1-h}}$ is the free energy of a
helix beginning with base pair $\Rbp{k}{j+1}$ and ending with base
pair $\Rbp{k+h}{j+1-h}$ according to \eq~\eqref{eq:5}. $\RGl(m+2)$ is
the free energy of a loop consisting of $m+2$ segments as given by
\eq~\eqref{eq:1}. $\RGl$ and $\RGh$ contain all interactions discussed
in the previous section.
\Eq~\eqref{eq:11} allows to compute the partition function in
polynomial time ($\order(\RN^4)$). Further, our formulation allows to
treat non-linear functions for $\RGl(\Rm)$ and $\RGh(h)$; for
instance, $\RGl(\Rm)$ is strongly non-linear by virtue of
\eqs~\eqref{eq:2} and~\eqref{eq:3}.

The unrestricted partition function of the entire RNA, where the
number of non-nested backbone bonds $\RM$  is allowed to fluctuate, is given
by
\begin{equation}
  \RQG_{\RN} = \sum_{\RM = 0}^N \exp[-\beta \RgfFJC \RM]\RQHt_{0,N}^{\RM}
  \label{eq:12}
\end{equation}
and contains the influence of force \via $\RgfFJC$ defined in
\eq~\eqref{eq:9}.
The partition function $\RQG_{\RN}$ contains all secondary structure
interactions, but neglects pseudoknots and other tertiary
interactions.  As has been argued before, this approximation is known
to work very well~\cite{Tinoco1999} and yields reliable predictions
for the stability of nucleic acids~\cite{Gruber2008}.

Using the same ideas, we determine the minimum free energy (mfe) and
the mfe structure.  The mfe structure, is defined as the secondary
structure, which gives the largest contribution to the partition
function. Since it cannot be derived from the partition function
itself, it has to be determined from a slightly modified set of
recursion relations, see supporting material.

\section{Salt dependence of melting curves}
\label{sec:salt-depend-melt}

\begin{figure}
  \centering
  \includegraphics{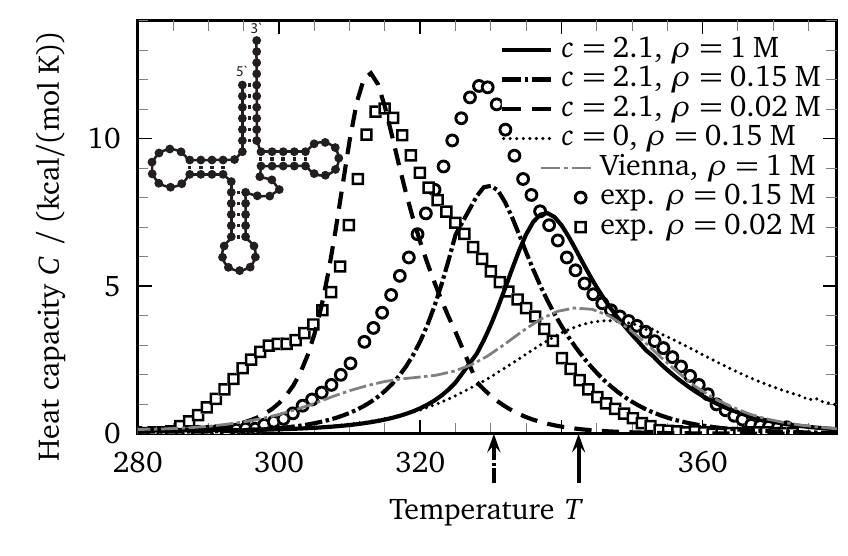}%
  \caption{ Melting curve of the 76 bases long tRNA-phe of yeast; the minimum free energy
    structure at $\con = \unit{1}{\molar}$, $\RT = \unit{300}{\kelvin}$, $\Rc = 2.1$ is shown as an inset.
    Symbols denote experimental melting curves for NaCl concentrations $\con =
    \unit{20}{\milli\molar}$ (squares) and $\unit{150}{\milli\molar}$ (circles)~\cite{Privalov1978}.
Our predictions for different salt concentrations are depicted by
the dashed ($\unit{20}{\milli\molar}$), dash-dotted
($\unit{150}{\milli\molar}$), and solid ($\unit{1}{\molar}$)
lines. The respective arrows indicate melting temperatures obtained
by experiments of another group~\cite{Vives2002} at the same salt
concentration $\con=\unit{150}{\milli\molar}$ (left arrow) and
$\con=\unit{1}{\molar}$ (right arrow). The dotted line shows our
prediction for $\con = \unit{150}{\milli\molar}$ and $\Rc = 0$ and
exemplifies that a non-zero loop exponent is responsible for
 rendering the transition more cooperative, in closer agreement with
  experiment; for $\unit{150}{\milli\molar}$ and $\Rc = 0$ the
melting temperature is at higher temperatures since the energy
parameters are optimized for $\Rc = 2.1$. The gray dash-dotted curve
is the prediction of the Vienna package, which uses a linearized
multi-loop entropy corresponding to $\Rc = 0$ and $\con=
\unit{1}{\molar}$. This is to be compared to our prediction for $\Rc
= 2.1$ and $\con = \unit{1}{\molar}$: while the melting temperatures
are similar, the cooperativity, \ie the widths of the peaks are
different due to different loop exponents. }
\label{fig:3}
\end{figure}
In this section we calculate melting curves for different salt
concentrations by applying \eqs~\eqref{eq:1} and~\eqref{eq:5}, which
include our salt dependent free energy parameterization. In
\fig~\ref{fig:3} we compare experimental
results~\cite{Privalov1978,Vives2002} with our predictions for the
heat capacity of yeast tRNA-phe; the sequence is given in the
supporting material section~D.
The heat capacity is readily obtained by
\begin{equation}
  \Rheatcap = \RT \frac{\partial^2\kBT \ln \RQG_N}{\partial\RT^2}
  \eqspace,\label{eq:13}
\end{equation}
where $\RQG_\RN$ is the unrestricted partition function of the RNA at
zero force, \eq~\eqref{eq:12}.  In all our calculations, we use the
same literature parameter set for the stacking and pairing free energy
$\Rghstack$. No additional fit parameter enters except the loop
initialization free energy $\RGlinit$, which is determined in
\fig~\ref{fig:2}a from a separate experimental data set. The salt
dependence of the experimentally observed melting temperatures is
reproduced well, compare \fig~\ref{fig:3}.  The arrows indicate
additional experimental results~\cite{Vives2002} for the melting
temperature for $\con = \unit{150}{\milli\molar}$ and
$\unit{1}{\molar}$, which again coincide with our prediction. We also
plot a calculated melting curve for loop exponent $\Rc = 0$ and NaCl
concentration $\con = \unit{150}{\milli\molar}$, which exhibits a far
less cooperative transition than observed in the corresponding curve
with $\Rc = 2.1$. Finally, we compare our prediction for $\con =
\unit{1}{\molar}$ and $\Rc = 2.1$ with the prediction of RNAheat in
the Vienna Package~\cite{Hofacker1994} for $\con = {1}{\molar}$, which
uses a linearized multi-loop entropy amounting to $\Rc = 0$ in our
framework. The predicted melting temperatures are almost
identical. However, the widths of the peaks in both melting curves
differ and our  melting profile for $\Rc = 2.1$ is more peaked.  Taking
all these observations together leads to the conclusion that only a
combined use of logarithmic loop entropy (characterized by a non-zero
loop exponent) and salt dependent free energy corrections leads to a
correct prediction of melting curves.  The additional features in the
experimental data, \eg the shoulder at lower temperatures and the
increased width of the experimental curves might be attributed to
tertiary structure rearrangements, which are not captured by our
approach, or to melting occurring in multiple stages.

\section{Salt dependence of stretching curves}
\label{sec:salt-depend-stretch}
\begin{figure}
  \centering
  \includegraphics{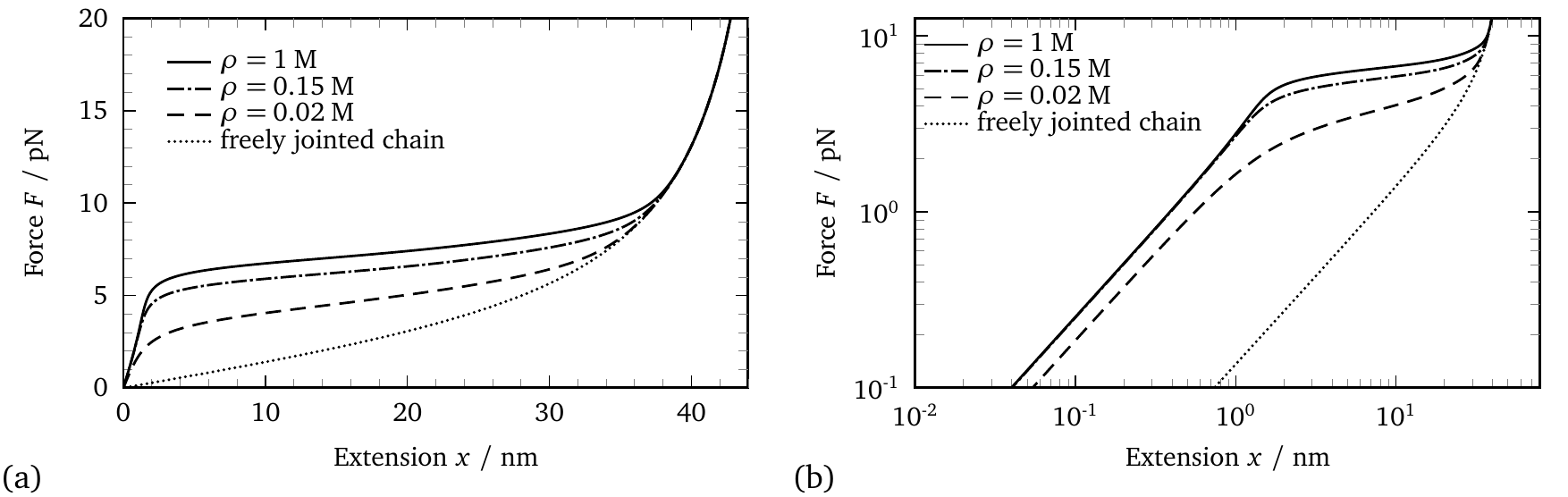}
  \subfloat{\label{fig:4a}}%
  \subfloat{\label{fig:4b}}%
  \caption{Salt dependence of stretching curves of tRNA-phe for
  different salt concentrations $\con =
  \unit{20}{\milli\molar},\,\unit{150}{\milli\molar},\,\unit{1}{\molar}$. Increasing
  salt concentration stabilizes the secondary structure due to
  screening of the electrostatic interaction. The dotted line is the
  theoretical prediction for the force extension curve of a freely
  jointed chain, \eq~\eqref{eq:10}. The deviation of the predicted
  curves for RNA from the FJC curve is due to the presence of
  secondary structure. The observed plateau force is due to the
  rupture of the secondary structure. We show the force extension
  curves in \subref{fig:4}a~a linear plot and \subref{fig:4}b~a
  double logarithmic plot, indicating that the force extension curve
  is linear in the low-force regime, before the secondary structure
  is ruptured apart.}
  \label{fig:4}
\end{figure}

Apart from temperature, force is an important variable to study
denaturation of RNA
molecules~\cite{Li2008,Liphardt2001,Manosas2006,Seol2007,Tinoco2004,Woodside2006,Woodside2006a,Hyeon2005,Gerland2001,Kumar2008a,Mueller2002,Montanari2001}.
In \fig~\ref{fig:4} we show the salt dependence of stretching curves
for yeast tRNA-phe. The stretching curves have been obtained by
describing the force response of the $\RM$ non-nested backbone bonds,
see \fig~\ref{fig:1}, with the freely jointed chain (FJC) model, see
\eq~\eqref{eq:10},
\begin{equation}
  \label{eq:14}
  x(\RF) =  \kBT \frac{\partial \ln \RQG_N}{\partial\RF} = \kBT
  \frac{\partial\ln\RQG_N}{\partial\RgfFJC}\frac{\partial\RgfFJC}{\partial\RF} = M x^{\mathrm{FJC}}(\RF)
  \eqspace,
\end{equation}
where we used the expectation value of the number of non-nested backbone
segments
\begin{equation}
  \label{eq:15}
  \RM = -\kBT \frac{\partial\ln\RQG_N}{\partial\RgfFJC}
  \eqspace.
\end{equation}

\begin{figure}
  \centering
  \includegraphics{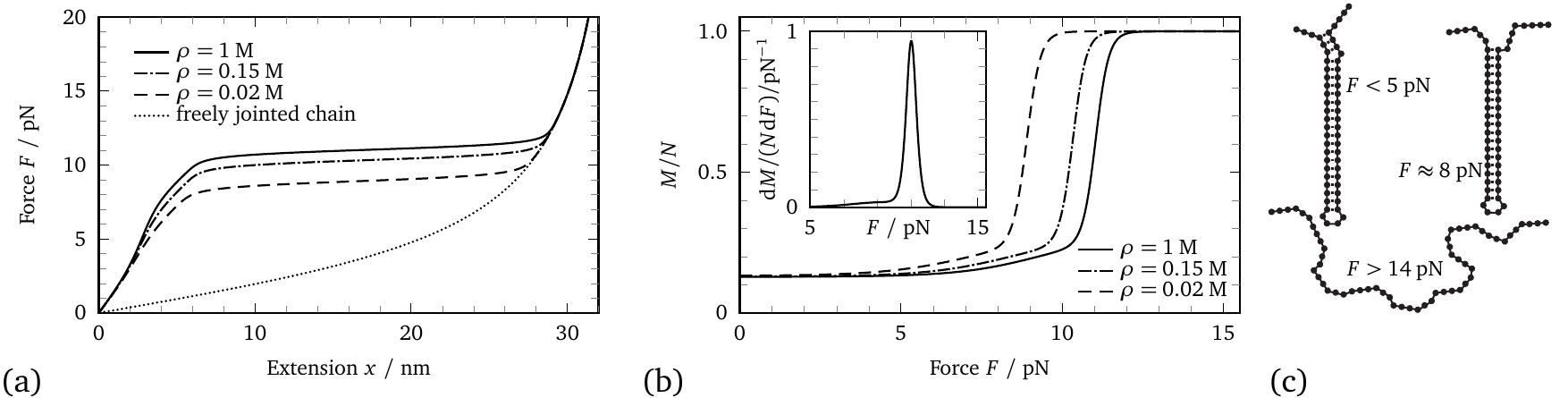}%
  \subfloat{\label{fig:5a}}%
  \subfloat{\label{fig:5b}}%
  \subfloat{\label{fig:5c}}%
  \caption{\subref{fig:5}a~Salt dependence of stretching curves of
  the 56~bases long RNA hairpin P5ab~\cite{Liphardt2001} for
  different salt concentrations $\con =
  \unit{20}{\milli\molar},\,\unit{150}{\milli\molar},\,\unit{1}{\molar}$. Increasing
  salt concentration stabilizes the secondary structure due to
  screening of the electrostatic interaction. The dotted line is the
  force extension curve of an FJC,
  \eq~\eqref{eq:10}. \subref{fig:5}b~The fraction of non-nested
  segments $\RM/\RN$ as a function of force. One observes that the
  unzipping of the hairpin occurs in two stages, which is visible as
  a shoulder for $\unit{5}{\pico\newton}\lesssim\RF\lesssim
  \unit{6-10}{\pico\newton}$ (exact values depend on the salt
  concentration) and a successive cooperative transition. The inset
  shows the derivative $\drm\RM/(\RN\drm\RF)$ for $\con =
  \unit{1}{\molar}$, where the first transition is visible as a
  shoulder at $\RF\approx\unit{8}{\pico\newton}$. The sharp peak at
  $\RF=\unit{11}{\pico\newton}$ is the rupture of the complete
  helix. \subref{fig:5}c~Predicted minimum free energy structures of
  the hairpin P5ab at different forces, see also the supporting
  material
  section~A. For $\RF
  < \unit{5}{\pico\newton}$ the hairpin is in the native state with
  all base pairs intact.  At $\RF\approx\unit{8}{\pico\newton}$ the
  first helix, consisting of three base pairs and bounded by the
  bulge loop, is ruptured. This causes the first smooth
  transition. Forces $\RF \gtrsim \unit{14}{\pico\newton}$ lead to
  the unzipping of the whole hairpin in a very cooperative fashion.}
  \label{fig:5}
\end{figure}

\begin{figure}
  \centering
  \includegraphics{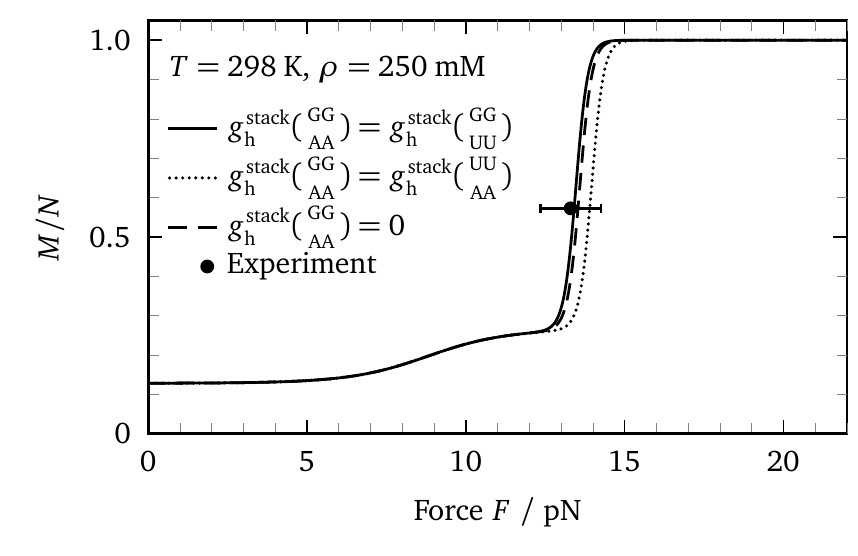}
  \caption{The effect of different parameterizations for the free
  energy parameters for ${{\mathrm{GG}}\atop{\mathrm{AA}}}$ on the
  denaturation curve is marginal.  Here, the fraction of non-nested
  backbone bonds is plotted against the force for the P5ab hairpin
  and $\RT = \unit{298}{\kelvin}$, $\con = {250}{\milli\molar}$,
  $\Rc = 2.1$. The solid line is obtained with the parameters used
  for the P5ab hairpin in the rest of this paper
  $\Rghstack({{\mathrm{GG}}\atop{\mathrm{AA}}}) =
  \Rghstack({{\mathrm{GG}}\atop{\mathrm{UU}}})$. The dotted line is
  obtained by using $\Rghstack({{\mathrm{GG}}\atop{\mathrm{AA}}}) =
  \Rghstack({{\mathrm{UU}}\atop{\mathrm{AA}}})$, whereas the dashed
  line is for $\Rghstack({{\mathrm{GG}}\atop{\mathrm{AA}}}) =
  0$. All three curves coincide and differ only slightly at the
  transition, exhibiting only marginally different transition
  forces, which all
 agree with the experimentally observed unfolding 
  force within errors\citet{Liphardt2001}. The values
 of the free energy parameters are given
  in the supporting material.}
  \label{fig:9}
\end{figure}

As for the melting curves, one observes that increasing salt
concentration stabilizes the structure, leading to higher unfolding
forces. All curves converge in the large force limit to a freely
jointed chain of the length of the whole RNA molecule $(N-1)\Rlss$,
where $N = 76$ is the number of bases in the chain. The deviation for
small forces from this theoretical prediction is due to the secondary
structure of RNA, which is present at small forces and which becomes
disrupted at forces $\RF \gtrsim \unit{\text{3-7}}{\pico\newton}$.  In
\fig~\ref{fig:5}a we show the force extension curve of the P5ab
hairpin~\cite{Liphardt2001}; the sequence is given in the supporting
material section~D. Apart from the
salt dependence of the force extension curve, one observes that the
unzipping of the helix occurs in two stages. This is seen best by
considering the fraction of non-nested segments and its derivative,
\fig~\ref{fig:5}b. The first stage is a smooth unzipping of the first
three base pairs up to the bulge loop visible as a shoulder at
$\RF\approx{8}{\pico\newton}$ in the derivative.  The second stage is
a sharp transition, where the rest of the hairpin unzips.  In
\fig~\ref{fig:5}c we show mfe predictions for the secondary structure
at different forces for $\con = \unit{1}{\molar}$ NaCl. For $\RF <
\unit{5}{\pico\newton}$, we predict correctly the experimentally
observed native state with all base pairs
intact~\cite{Liphardt2001}. For forces
$\RF\approx\unit{8}{\pico\newton}$, an intermediate state appears,
where the first three base pairs are unzipped up to the bulge
loop. Denaturation is observed for $\RF \gtrsim
\unit{14}{\pico\newton}$.
The native structure of the P5ab hairpin contains the stacked pairs
${{\mathrm{GG}}\atop{\mathrm{AA}}}$ --~bp\Rbp{17}{42} and
bp\Rbp{18}{41}~\cite{Liphardt2001}. For this stack, no free energy
parameters are available and we use the parameters for the stack
${{\mathrm{GG}}\atop{\mathrm{UU}}}$, instead. However, other
parameterizations for this stack work equally well and reproduce the
experimental transition force within errors, see \fig~\ref{fig:9}.

\section{Phase diagrams of RNA hairpin P5ab}
\begin{figure}
  \centering
  \subfloat{\label{fig:6a}}%
  \subfloat{\label{fig:6b}}%
  \includegraphics{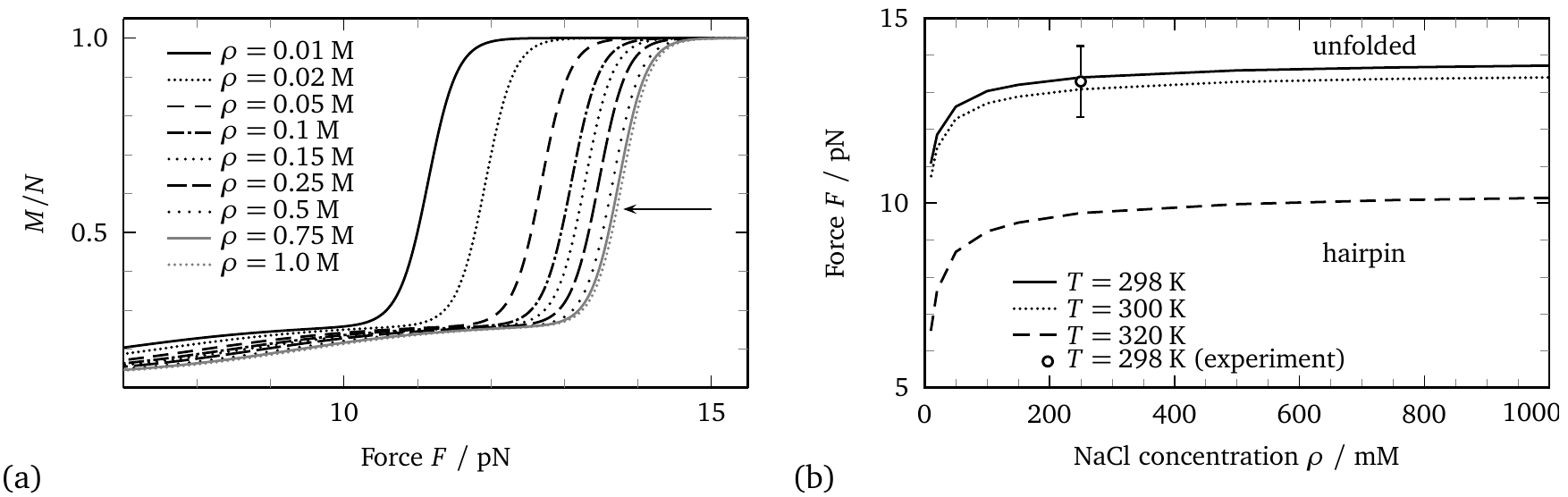}%
  \caption{\subref{fig:6}a~Fraction of non-nested segments of the
  P5ab hairpin as a function of force for different salt
  concentrations and constant temperature $\RT =
  \unit{298}{\kelvin}$. The position of the crossover, which is
  defined as the point where $\RM - 7 = (\RN-7)/2$, \ie
  $\RM/\RN=0.56$ (indicated by the arrow), determines the phase
  diagram. \subref{fig:6}b~Phase diagram of the P5ab hairpin in the
  $\RF$-$\con$ plane for different temperatures $\RT =
  \unit{298}{\kelvin},\,\unit{300}{\kelvin},\,\unit{320}{\kelvin}$. Below
  the curve the RNA is in the hairpin phase, above the RNA is
  denatured. The symbol at $\con = \unit{250}{\milli\molar}$,
  $\RF=\unit{13.3}{\pico\newton}$, and $\RT = \unit{298}{\kelvin}$
  denotes the experimental data by \citet{Liphardt2001} and
  coincides with our prediction.}
  \label{fig:6}
\end{figure}

\begin{figure}
  \centering
  \includegraphics{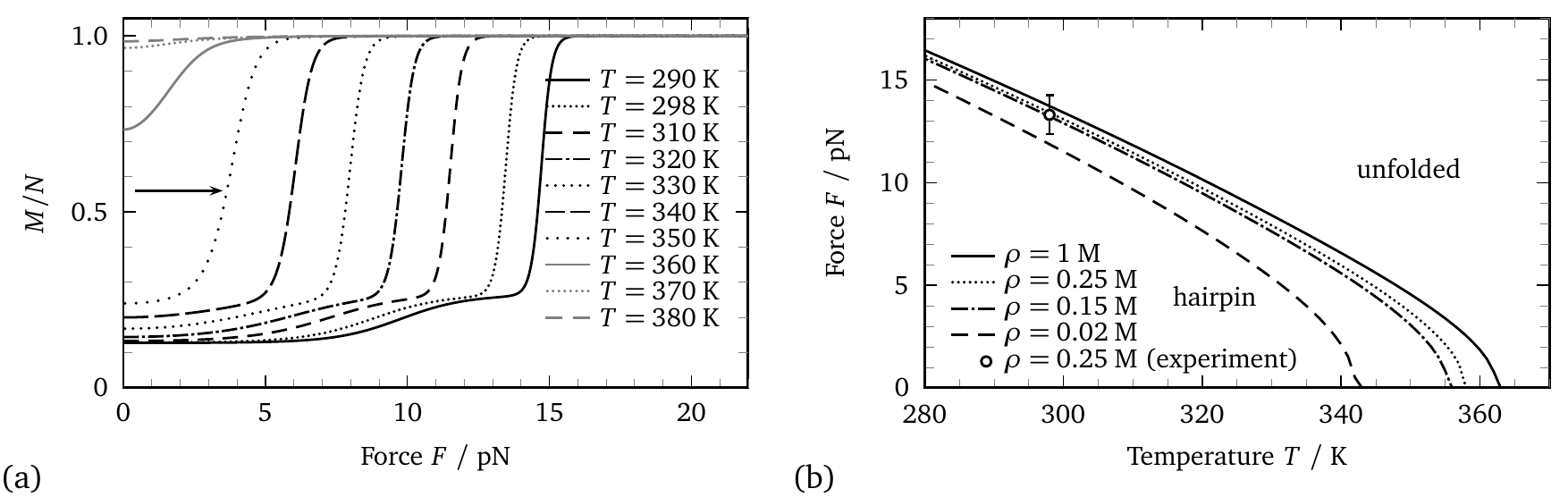}%
  \subfloat{\label{fig:7a}}%
  \subfloat{\label{fig:7b}}%
  \caption{\subref{fig:7}a~Fraction of non-nested segments of the
  P5ab hairpin as a function of force for different temperatures and
  constant salt concentration $\con = \unit{250}{\milli\molar}$. The
  position of the crossover (arrow, $\RM/\RN = 0.56$) determines the
  phase boundary. With increasing temperature a decrease of the
  denaturation force is observed. Above the melting temperature
  $\RTmelt\approx\unit{358}{\kelvin}$ the molecule is always in the
  denatured state. \subref{fig:7}b~Phase diagram of the P5ab hairpin
  in the $\RF$-$\RT$ plane. Below the curve the RNA is in the native
  hairpin phase, above the RNA is denatured. The symbol denotes
  experimental values~\cite{Liphardt2001}.}
  \label{fig:7}
\end{figure}

\begin{figure}
  \centering
  \includegraphics{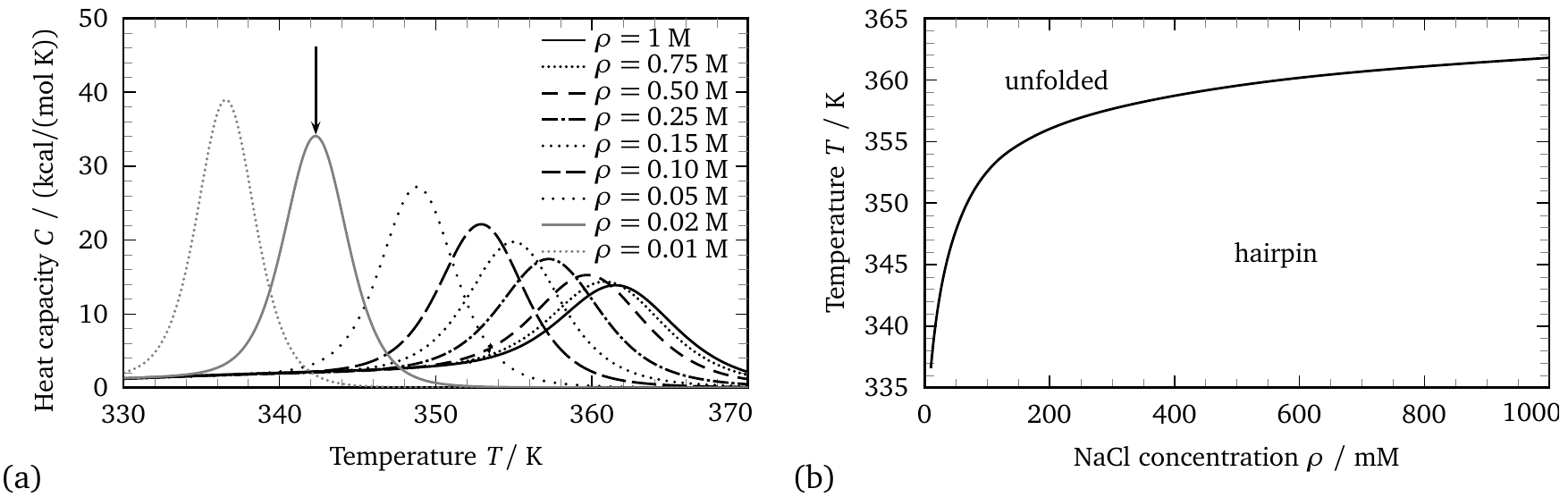}%
  \subfloat{\label{fig:8a}}%
  \subfloat{\label{fig:8b}}%
  \caption{\subref{fig:8}a~Heat capacity curves for different salt
 concentrations as a function of temperature. The peak position
  moves to higher temperatures with increasing salt concentration. The positions
  of the peaks, denoted exemplarily for the curve with
  $\con=\unit{20}{\milli\molar}$ by the arrow, determine the phase
  diagram. \subref{fig:8}b~Phase diagram of the P5ab hairpin in the
  $\RT$-$\con$ plane. Below the curve the RNA is in the native
  hairpin phase, above the RNA is denatured.}
  \label{fig:8}
\end{figure}

With the tools established in the previous sections, we are now able
to study phase diagrams of RNA. We consider the P5ab hairpin, which is
a well studied
system~\cite{Liphardt2001,Gerland2003,Hyeon2005,Wen2007,Cocco2003}.
In \fig~\ref{fig:6}b the phase diagram in the $\RF$-$\con$ plane is
shown for $\RT =
\unit{298}{\kelvin},\,\unit{300}{\kelvin},\,\unit{320}{\kelvin}$ and
$\Rc = 2.1$. The phase boundary is defined as the force where half of
the helical section is unzipped. For the definition of the phase
boundary, we exclude the three unpaired bases at the 5'- and the four
bases at the 3'-end, see \fig~\ref{fig:5}c, and use the condition $\RM
- 7 = (\RN-7)/2$. This threshold value of $\RM/\RN$ is depicted by an
arrow in \fig~\ref{fig:6}a. Below the phase boundary, the hairpin is
stable, above the molecule is denatured.  In \fig~\ref{fig:8}b we
additionally include the experimental results by \citet{Liphardt2001}
agreeing nicely with our results.  It is important to
note, that this transition is not a phase transition in the strict
statistical mechanics sense, but just a crossover. A true phase
transition is defined as a non-analyticity of the free energy, which
can only occur for an infinite system with long-range
interactions~\cite{Einert2008}.  The three-dimensional phase space we
are considering is spanned by temperature, force, and salt
concentration. In \figs~\ref{fig:7} and \ref{fig:8} we show slices in
the $\RF$-$\RT$ and in the $\RT$-$\con$ plane. The phase boundary for
the $\RF$-$\RT$ plane is determined the same way as in the
$\RF$-$\con$ plane, yet with varying temperature and fixed salt
concentration. The phase boundary in the $\RT$-$\con$ plane is
determined differently: heat capacity curves as a function of
temperature are calculated for different salt concentrations.  The
position of the peaks in the heat capacity curves (one is depicted by
an arrow in \fig~\ref{fig:8}a) determine the phase diagram in
\fig~\ref{fig:8}b. Therefore, slight differences between the phase
diagrams in \figs~\ref{fig:6},~\ref{fig:7} on the one hand and
\fig~\ref{fig:8} on the other hand may arise.

We observe that for large salt concentrations, the denaturation forces
and temperatures are rather independent of the salt concentration, see
\figs~\ref{fig:6} and~\ref{fig:8}. Only when the Debye screening
length $\Rkappa^{-1}$ is of the order of the typical length scale of
RNA, which is the case for $\con\lesssim\unit{100}{\milli\molar}$, a
marked dependence on the salt concentration is observed.

\section{Conclusions}
\label{sec:conclusions}
We construct a theory for RNA folding and melting that includes the
effects of monovalent salt, loop entropy, and stretching forces.  Our
theory is based on salt and temperature dependent modifications of the
free energies of RNA helices and loops that include electrostatic
interactions on the linear Debye-H\"uckel level --~augmented by
Manning condensation~-- and conformational fluctuation effects \via
the asymptotic, non-linear expression for the entropy of loop
formation.  Decreasing salt concentration is shown to generally
destabilize RNA folds and to lower denaturation temperatures and
forces.  The predictions are in 
good  agreement with experimental
data as shown for two different scenarios, namely the heat capacity
curves for the thermal denaturation of tRNA-phe and the response of
the P5ab RNA hairpin to an external pulling force.

Due to the usage of the linear Debye-H\"uckel approximation in
conjunction with the Manning condensation concept, our approach is
limited to monovalent salt and neglects ion-specific effects.
Electrostatic nonlinear and correlation effects could in principle be
taken into account by more advanced modeling using variational
approaches~\cite{Netz2003}, while ion-specific effects could be
straightforwardly included using effective interactions between
different ions and RNA bases~\cite{Schwierz2010}.  More complex
phenomena involving multivalent ions such as Mg\textsuperscript{2+}
could in principle be modeled by allowing for a few tertiary contacts,
which is left for future studies.

We find that for a proper description of RNA melting curves, correct
modeling of the loop entropy is crucial.  A non-zero loop exponent
leads to an increased cooperativity of the melting transition and thus
makes the heat capacity curve narrower in good agreement with
experimental results.  We conclude that for a correct description of
RNA denaturation thermodynamics, both loop entropy and salt effects
are important and should be included in standard structure and melting
curve prediction software.

\section{Acknowledgements}
\label{sec:acknowledgements}

Financial support comes from the DFG \via grant NE
810/7. T.R.E. acknowledges support from the Elitenetzwerk Bayern
within the framework of CompInt.

\bibliography{RNA_salt}

\clearpage



\section*{Supplementary material}

# Supporting material
# Theory for RNA folding, stretching, and melting including loops and salt

Thomas R. Einert[1, *] and Roland R. Netz[1, †]

[1]*Physik Department, Technische Universität München, 85748 Garching, Germany*

(Dated: April 19, 2011)

The supporting material provides details on how to obtain the minimum free energy structure by a recursive backtracking algorithm. Interpolation formulas for the salt parameterization of the free energy parameters are given. Information on the binding free energy parameters as well as the sequences used is given.

### Contents

## A.  MINIMUM FREE ENERGY STRUCTURE PREDICTION

The partition function involves the enumeration and contribution of all possible secondary structures and can be calculated exactly by the recursion relation eq. (11). From the partition function $Q_{i,j}^M$ the free energy of an RNA strand ranging from base $i$ through $j$ with $M$ non-nested backbone bonds can be calculated

$$\mathcal{F}_{i,j}^M = -\mathrm{k_B} T \ln Q_{i,j}^M \,. \tag{S1}$$

Another important quantity is the minimum free energy $\mathcal{E}$ (mfe) associated with the ground state structure, also called mfe structure. In analogy to the partition function we define the mfe of a substrand ranging from base $i$ through $j$ with $M$ non-nested backbone segments as $\mathcal{E}_{i,j}^M$. RNA structure prediction relies on the assumption that the native – experimentally observed – structure is given by the mfe structure. Following the same idea of the recursive calculation of the partition function, see eq. (11), we can determine the minimum free energy of a sequence by replacing the sums by the min operator and obtain the recursion relations

$$\mathcal{E}_{i,j+1}^{M+1} = \min\left\{ \mathcal{E}_{i,j}^M, \min_{k=i+1,\ldots,j}\left\{ \mathcal{E}_{i,k-1}^M + \mathcal{E}_{k,j+1}^0 \right\} \right\} \tag{S2a}$$

$$\mathcal{E}_{k,j+1}^0 = \min_{m=0,\ldots,j+1-k}\left\{ \mathcal{E}_{k+1,j}^m + \varepsilon_{k,j+1} \right\} \,. \tag{S2b}$$

$\varepsilon_{k,j+1}$ is the free energy of the base pair $(k, j+1)$. From the three-dimensional array the mfe structure can now be obtained by a recursive backtracking algorithm. The following pseudo code initiates the backtracking

```
 E_buffer = ∞;
 for( M = 0; M < N; M++ ){
   if( E_{0,N}^M  <  E_buffer ){
     E_buffer = E_{0,N}^M;
```

---

*Corresponding author: Physik Department, Technische Universität München, James-Franck-Straße, 85748 Garching, Germany, Tel.: +49-89-28914337, Fax: +49-89-28914642, E-mail: einert@ph.tum.de

†E-mail: netz@ph.tum.de

```
    M_min = M;
  }
}

backtrack(0, N, M_min);
```

The function `backtrack(i, j, M)` performs the next backtracking step and calls itself recursivly. Its pseudocode reads

```
backtrack(i, j, M){
  /* Check if (i, j, M) is an allowed triple, this includes e.g.
        -    M ≤ j-i
        -    i≤j
        -    if M == 0 then base i and base j must be complemetary
        -    ...
     If (i, j, M) is not allowed, then  E_{i,j}^M  == ∞ */
  if( E_{i,j}^M  < ∞){
    if( M == 0 ){
      /* i.e. base i and j are paired => add to list of base pairs */
      add_to_list_of_base_pairs(i, j);
      E_buffer = ∞;
      /* Determine the size m_min of the loop closed by the pair (i,j) */
      for( m = 0; m ≤ j-i; m++){
        if( E_{i+1,j-1}^m  < E_buffer ){
          E_buffer = E_{i+1,j-1}^m ;
          m_min = m;
        }
      }
      /* Perform the next backtracking step */
      backtrack(i+1, j-1, m_min);
    }
    else if( M > 0 ){
      E_buffer = ∞;
      for( k = 0; k ≤ j; k++ ){
        /* Split the structure into two structures. The left structure
           ranges from i through k-1 and has M-1 non-nested backbone bonds.
           The right structure ranges from k through j and has zero
           non-nested backbone bonds, i.e. it is either a single
           base (k==j) or is a closed structure terminated by the pair (k,j) */
        if( E_{i,k-1}^{M-1} + E_{k,j}^0 < E_buffer ){
          E_buffer = E_{i,k-1}^{M-1} + E_{k,j}^0 ;
          k_min = k;
        }
      }
      /* Perform next backtracking step on the left substructure */
      backtrack(i, k-1, M-1);
      /* Perform next backtracking step on the right substructure */
      if( k ≠ j ){
        backtrack(k, j, 0);
      }
    }
  }
}
```

## B. INTERPOLATION FORMULAS

In order to make our results more user-friendly we give interpolation formulas for the salt corrections which involve hypergeometric and Bessel functions as those might not be available for every user.

The salt correction for the free energy of loops contains two generalized hypergeometric functions

$$\mathcal{G}_l^{\text{salt}} = k_B T l_B (m l_{ss}) \tau_{ss}^2 \left[ \ln(\kappa m l_{ss}) - \ln(\pi/2) + \gamma - \frac{\kappa m l_{ss}}{2} {}_1F_2 \left( 1/2, \binom{1}{3/2}, \left( \frac{\kappa m l_{ss}}{2\pi} \right)^2 \right) \right.$$
$$\left. + \frac{1}{2} \left( \frac{\kappa m l_{ss}}{\pi} \right)^2 {}_2F_3 \left( \binom{1}{1}, \binom{3/2}{3/2}{2}, \left( \frac{\kappa m l_{ss}}{2\pi} \right)^2 \right) + \frac{1}{\kappa m l_{ss}} \left( 1 - \exp(-\kappa m l_{ss}) + \kappa m l_{ss} \Gamma(0, \kappa m l_{ss}) \right) \right] . \tag{S3}$$

The sum of the two hypergeometric functions is well approximated by interpolating between the two asymptotic expansions for small and large argument

$$- \frac{y}{2} {}_1F_2 \left( 1/2, \binom{1}{3/2}, \left( \frac{y}{2\pi} \right)^2 \right) + \frac{1}{2} \left( \frac{y}{\pi} \right)^2 {}_2F_3 \left( \binom{1}{1}, \binom{3/2}{3/2}{2}, \left( \frac{y}{2\pi} \right)^2 \right)$$
$$\approx \frac{1}{\frac{y^6}{(2\pi)^6} + 1} \left( \frac{y^4}{36\pi^4} - \frac{y^3}{24\pi^2} + \frac{y^2}{2\pi^2} - \frac{y}{2} \right) + \left( 1 - \frac{1}{\frac{y^6}{(2\pi)^6} + 1} \right) \left( \log \left( \frac{2\pi}{y} \right) - 1.96351 \right) . \tag{S4}$$

The salt correction of the binding free energy of a base pair contains a modified Bessel function of the second kind, eq. (8),

$$\mathcal{G}_h^{\text{salt}} / h = g_h^{\text{DH}}(\rho) - g_h^{\text{DH}}(1 \text{ M}) , \tag{S5}$$

with

$$g_h^{\text{DH}}(\rho) = 2 k_B T \tau_{ds}^2 l_{ds} l_B K_0(\kappa d) . \tag{S6}$$

Eq. (S5) is well approximated by the heuristic formula

$$\mathcal{G}_h^{\text{salt}} / h \approx k_B T \tau_{ds}^2 l_{ds} l_B \left( b_1 + \frac{b_2 - b_3 \ln(b_4 \kappa)}{1 + b_5 \kappa} \right) , \tag{S7}$$

with the constants $b_1 = 0.0315171$, $b_2 = 35.0754$, $b_3 = 1.62292$, $b_4 = 1$ nm, $b_5 = 4.26381$ nm.

## C. FREE ENERGY PARAMETERS

The enthalpy $h$ and entropy $s$ parameters used in our calculations are taken from reference [1, 2]. For instance, the entries in the row UA and the column GC in the following table, $h_{\text{UA,GC}}$ and $s_{\text{UA,GC}}$, give the enthalpy and entropy contribution due to the stacking of the two neighboring base pairs UA and GC, where U and C are located at the 5'-end. The two bottom rows contain the initiation and termination contribution to helices and loops. For instance, the total free enthalpy of the triple helix terminated by a hairpin of length $m = 6$, $\substack{5'-\text{CGA}-\\3'-\text{GCU}-}loop$, at 1 M NaCl is given by

$$\begin{aligned}\mathcal{G} &= \mathcal{G}_{\text{h}}^{\text{init}} + \mathcal{G}_{\text{h}}^{\text{stack}} + \mathcal{G}_{\text{h}}^{\text{term}} + \mathcal{G}_{\text{l}}^{\text{init}} + \mathcal{G}_{\text{l}}^{\text{conf}} \\ &= h_{\text{h}}^{\text{init}} - Ts_{\text{h}}^{\text{init}} \\ &\quad + h_{\text{CG,GC}} + h_{\text{GC,AU}} - T(s_{\text{CG,GC}} + s_{\text{GC,AU}}) \\ &\quad + h_{\text{h}}^{\text{term}} - Ts_{\text{h}}^{\text{term}} \\ &\quad + h_{\text{l}}^{\text{init}} - Ts_{\text{l}}^{\text{init}} \\ &\quad - k_{\text{B}} T \ln m^{-c} \,.\end{aligned} \tag{S8}$$

| | Enthalpy $h$ / (kcal/mol) | | | | | | | Entropy $s$ / $(10^{-3}\text{kcal/(mol K)})$ | | | | | |
|---|---|---|---|---|---|---|---|---|---|---|---|---|---|
| | AU | UA | CG | GC | GU | UG | | AU | UA | CG | GC | GU | UG |
| AU | −6.82 | −9.38 | −11.40 | −10.48 | −3.21 | −8.81 | AU | −19.0 | −26.7 | −29.5 | −27.1 | −8.6 | −24.0 |
| UA | −7.69 | −6.82 | −12.44 | −10.44 | −6.99 | −12.83 | UA | −20.5 | −19.0 | −32.5 | −26.9 | −19.3 | −37.3 |
| CG | −10.44 | −10.48 | −13.39 | −10.64 | −5.61 | −12.11 | CG | −26.9 | −27.1 | −32.7 | −26.7 | −13.5 | −32.2 |
| GC | −12.40 | −11.40 | −14.88 | −13.39 | −8.33 | −12.59 | GC | −32.5 | −29.5 | −36.9 | −32.7 | −21.9 | −32.5 |
| GU | −12.83 | −8.81 | −12.59 | −12.11 | −13.47 | −14.59 | GU | −37.3 | −24.0 | −32.5 | −32.2 | −44.9 | −51.2 |
| UG | −6.99 | −3.21 | −8.33 | −5.61 | −9.26 | −13.47 | UG | −19.3 | −8.6 | −21.9 | −13.5 | −30.8 | −44.9 |
| $h_{\text{h}}^{\text{init/term}}$ | 3.72 | 3.72 | 0.00 | 0.00 | 3.72 | 3.72 | $s_{\text{h}}^{\text{init/term}}$ | 10.5 | 10.5 | 0.0 | 0.0 | 10.5 | 10.5 |
| $h_{\text{l}}^{\text{init}}$ | 1.68 | 1.68 | 1.68 | 1.68 | 1.68 | 1.68 | $s_{\text{l}}^{\text{init}}$ | −0.7 | −0.7 | −0.7 | −0.7 | −0.7 | −0.7 |

## D. SEQUENCES

The sequence of yeast tRNA-phe reads [3]
GCGGAUUUAG CUCAGUUGGG AGAGCGCCAG ACUGAAGAUC UGGAGGUCCU GUGUUCGAUC CACAGAAUUC GCACCA.

The sequence of the P5ab hairpin reads [4]
ACAGCCGUUC AGUACCAAGU CUCAGGGGAA ACUUUGAGAU GGGGUGCUGA CGGACA.

---



\end{document}